\documentclass[aps,prl,showpacs,twocolumn]{revtex4}
\usepackage{amsfonts}
\usepackage{mathrsfs}
\begin{document}
\title{Reply to the Comment on ``Spin and orbital angular momentum
in gauge theories:
Nucleon spin structure and multipole radiation revisited''
[PRL 100:232002 (2008)]}
\author{Xiang-Song Chen,$^{1,2}$ Xiao-Fu L\"{u},$^1$ Wei-Min Sun,$^2$
Fan Wang,$^2$ and T. Goldman$^3$}
\affiliation{$^1$Department of
Physics, Sichuan University, Chengdu 610064, China\\
$^2$Department of Physics, Nanjing University, CPNPC, Nanjing
210093, China\\
$^3$Theoretical Division, Los Alamos National Laboratory, Los
Alamos, NM 87545, USA}
\date{\today}

\maketitle

As the comment of Tiwari \cite{Tiwa08} reflects typical and probably
common misunderstandings, we feel it valuable to elaborate further
on these delicate and important issues:

(i) In the gauge coupling $\bar\psi\gamma^\mu A_\mu\psi$, the gauge
field $A_\mu$ play a dual role: it provides a physical coupling to
the Dirac field $\psi$, as well as a gauge freedom to compensate for
the phase freedom of $\psi$. Our idea of solving the
gauge-invariance problem is to decompose this dual role by seeking,
in any gauge, a unique separation $\vec A=\vec A_{\rm pure}+\vec
A_{\rm phys}$, with $\vec A_{\rm pure}$ a pure-gauge term
transforming in the same manner as does the full $\vec A$ and always
giving null $\vec B$, and $\vec A_{\rm phys}$ a physical term
transforming in the same manner as does the electric field $\vec E$.
Namely, $\vec A_{\rm phys}$ is gauge invariant/covariant in
electrodynamics/Yang-Mills theory, while $\vec A_{\rm pure}$ has the
same gauge freedom as $\vec A$ and can be used instead of $\vec A$
to construct a covariant derivative $\vec D_{\rm pure}\equiv
\vec\nabla-i\vec A_{\rm pure}$. The separation $\vec A=\vec A_{\rm
pure}+\vec A_{\rm phys}$, and the definitions for $\vec A_{\rm
phys}$ and $\vec A_{\rm pure}$, are by no means specifying a gauge
or restricting the gauge freedom of $\vec A$. E.g., in
electrodynamics, $\vec \nabla\cdot \vec A_{\rm phys}=0$ is the
definition of $\vec A_{\rm phys}$ in any gauge, it is not to be
confused with the gauge condition $\vec \nabla\cdot \vec A=0$.
Nevertheless, the separation $\vec A=\vec A_{\rm pure}+\vec A_{\rm
phys}$ does become the simplest in a unique physical gauge in which
$\vec A_{\rm pure}=0$ and $\vec A_{\rm phys}= \vec A$. Coulomb gauge
$\vec \nabla \cdot \vec A=0$ is the physical gauge for
electrodynamics. But for Yang-Mills theory the physical gauge is
$[\vec A,\vec E]=0$ which makes the gauge-dependent gluon color
charge vanish \cite{Chen08}, while $\vec \nabla \cdot \vec A=0$ is
no longer privileged. This may elucidate why quantization of
Yang-Mills theory in the Coulomb gauge is not very illuminating.

(ii) Our solution achieves the best possible reconciliation of
Lorentz covariance and gauge invariance for the momentum and angular
momentum (AM) of coupled quarks and gluons. To appreciate this
point, we first note that, in the instant form, six Poincar\'e
generators can be interaction-free (or {\em good}): $\vec P=\vec
P_q+\vec P_g$, $\vec J=\vec J_q +\vec J_g$, and four generators
involve interactions: $H=H_q+H_g+H_{int}$, $\vec K=\vec K_q+\vec
K_g+\vec K_{int}$. Here $q$, $g$, and $int$ denote the quark, gluon,
and quark-gluon interacting parts, respectively. In gauge theories,
it should be further noted that not only the interaction-involving
Lorentz transformations, but also the interaction-free ones, are
troublesome. This is because an interaction-free operator, like
$\int d^3x \psi ^\dagger \vec x \times\frac 1i \vec \nabla \psi$, is
often gauge-dependent, while a gauge-invariant construction, like
$\int d^3x \psi ^\dagger \vec x \times\frac 1i \vec D \psi$, often
involves interaction. In fact, the essence of our contribution is
that the {\em good} generators $\vec P$ and $\vec J$ are indeed
constructed to be both interaction-free and gauge independent, so
that each part in $\vec P$ and $\vec J$ transforms properly under
spatial translations and rotations. But since $\vec K$ involve
interaction intrinsically, only the {\em total} $\vec J$ transform
properly under boost: $[K^i,J^j]=i\epsilon_{ijk} K^k$, while
$[K^i,J_{q,g}^j]=i\epsilon_{ijk} K_{q,g}^k$ can {\em not} hold
(otherwise we would have $[K^i,J_q^j+J_g^j]=i\epsilon_{ijk}
[K_q^k+K_g^k]$, which contradicts $[K^i,J^j]=i\epsilon_{ijk} K^k$).
To determine the boost properties of $\vec J_{q,g}$, we need to
carry out a canonical quantization (preferably in the physical gauge
so that the pure-gauge term vanishes), then compute the commutators
$[K^i,J_{q,g}^j]$. This task is difficult but unavoidable for
everyone. In comparison, the light-cone formalism renders boost
along the third axis interaction-free, but makes the description of
AM rather troublesome, because two AM components (namely, the
rotation generators along the $x$, $y$ axes) involve interactions.
All in all, the separation of $\vec P$ or $\vec J$ into quark and
gluon parts can never be fully covariant under all Lorentz
transformations, no matter how one deals with the gauge field. In
fact, this property is an intrinsic complication for any interacting
system, not merely gauge interactions. By decomposing $\vec A$ into
physical and pure-gauge terms, we add absolutely no extra
complication in this regard.

(iii) It is an illusion that the spin and orbital AM of light as in
the optic measurements or multipole-radiation analysis can be
straightforwardly interpreted in classical electrodynamics, without
applying the approach which we follow. As long as one discusses spin
and orbital AM separately, the gauge-invariance problem is sharply
encountered, no matter one adopts a classical or quantum language.
This problem is just absent if one merely concerns about the
integrated total AM. Without defining the spin and orbital AM
gauge-invariantly, it is impossible to interpret the optic
measurements which intend to manipulate spin and orbital AM
separately, or the multipole-radiation analysis which employs the
notion of spin-orbital coupling. We should also clarify that the
solution we provide applies to both classical and quantum dynamics.
It is our gauge-invariant construction for the spin and orbital AM
that justifies the spin and orbital values or quantum numbers
assigned to the electromagnetic field or photon in any experiment.
And in turn, the consistency of our theory with experimental results
supports the validity and correctness of our gauge-invariant
expressions.

(iv) To see the physical content of $\vec E\times\vec B$, note that
$\vec J=\int d^3x \vec x\times (\vec E\times\vec B)$ gives the total
AM of a free gauge field, including both spin and orbital parts. But
if $\vec P(x)$ is the momentum density, then $\int d^3x \vec x\times
\vec P(x)$ is the standard form of orbital AM. This implies that
$\vec E\times\vec B$ is not a pure mechanical momentum, it must
include a spin flow so as to give the total $\vec J$ by an apparent
orbital form. By writing $\vec J=\int d^3x \vec E \times \vec A_{\rm
phys}+ \int d^3x \vec x \times (E^i\vec \nabla A_{\rm phys}^i)$, the
correct momentum density is read out to be $E^i\vec \nabla A_{\rm
phys}^i$. From $\vec E\times \vec B=\vec E^i\vec \nabla A_{\rm
phys}^i -\nabla^i (E^i \vec A_{\rm phys})+(\vec \nabla \cdot \vec
E)\vec A_{\rm phys}$, we see that the difference between $\vec
E\times\vec B$ and $E^i\vec \nabla A^i_{\rm phys}$ is a surface term
only if the (in general charged, here) Dirac field is absent so that
$\vec \nabla \cdot \vec E=0$. For an interacting system, the
difference is substantial. Similar situation occurs for the AM:
$\vec x\times (\vec E\times\vec B)=\vec E \times \vec A_{\rm phys}+
\vec x \times (E^i\vec \nabla A_{\rm phys}^i)+ \nabla^i[E^i(\vec
A_{\rm phys}\times\vec x)]-(\vec \nabla\cdot \vec E)\vec A_{\rm
phys}\times \vec x$, hence $\int d^3x \vec x\times (\vec E\times\vec
B)$ may differ drastically from $\int d^3x \vec E \times \vec A_{\rm
phys}+ \int d^3x \vec x \times (E^i\vec \nabla A_{\rm phys}^i)$ in
the presence of interaction. For the nucleon, then, $\vec
E\times\vec B$ and $E^i\vec \nabla A^i_{\rm phys}$ may give totally
different views of how much nucleon momentum is carried by gluons.
$\vec E\times\vec B$ is a component of the symmetric energy-momentum
tensor, while $E^i\vec \nabla A^i_{\rm phys}$ belongs to the
(non-symmetric) canonical one. These two tensors differ by a
total-derivative spin term. This term does not matter for the
integrated total energy or momentum, so there is no curiosity in
that a photon with arbitrary polarization and orbital AM can have
the same energy. However, the spin term does alter the density and
symmetry of the energy-momentum tensor, for which we do need a
concrete density expression for the purpose of coupling to gravity.
Thus, were one to adopt Einstein's gravitational equation which
requires a symmetric tensor, then the ``momentum'' that contributes
to gravity is not the mechanical momentum \cite{Chen07}.

In closing, we remark that Ref. \cite{Chen08} does solve the
long-standing gauge-invariance problem of spin and orbital AM, in
the sense that the expressions satisfy all theoretical requirements
and agree with all experimental results.

\end{document}